\documentclass[onecolumn,amsmath,amssymb,pra,superscriptaddress]{revtex4}

\usepackage{graphics}
\usepackage{epsfig}
\usepackage{amsmath}
\usepackage{amsfonts}
\usepackage{color}
\usepackage{dcolumn}
\usepackage{bm}
\usepackage{mathrsfs}

\begin{document}

  \bibliographystyle{apsrev}
  
\title{Parameter-free Separable Lowest-order Non-relativistic Hamiltonian for Helium Atom}

      \author{E. O. Jobunga}
       \affiliation{\it  Department  of Mathematics and Physics , Technical University   of Mombasa,\\ P. O. Box 90420-80100, Mombasa, Kenya}
%
%


\begin{abstract}
Electron-electron correlation forms the basis of difficulties encountered in many-body physics. Accurate treatment of the correlation problem is likely to unravel some nice physical properties of matter embedded in the correlation. In an effort to tackle the many-body problem, an empirically determined symmetry-dependent partition fraction for the electron-electron interaction energy between two interacting states of helium atom is suggested in this study. Using the partition fraction and the lowest-order approximation of the multipole series expansion of the electron-electron interaction, a simple parameter-free pseudopotential for a two-electron system is derived. The groundstate, singly and doubly excited state non-relativistic energies generated by the pseudopotential are in reasonable agreement with literature values.

\end{abstract}

\maketitle
\section{Introduction}

Helium atom and helium-like ions are the simplest many-body systems containing two electrons which interact among themselves in addition to their interaction with the nucleus. The two-electron systems are therefore the ideal candidates for studying the  electron correlation effects. 

 Several theoretical approaches have been employed in the past in dealing with the electron correlation problem. Some of the approaches include the  variational Hylleraas method \cite{Hylleraas1929, Drake1999}, coupled channels method \cite{Barna2003}, the configuration interaction method \cite{Hasbani2000}, explicitly correlated basis and complex scaling method \cite{Scrinzi1998}. At present only the Hylleraas method, which includes the interelectronic distance as an additional free co-ordinate, yields the known absolute accuracy of the groundstate energy of the helium atom \cite{Pekeris1959}. 
 
 Configuration interaction methods have also been proved to be accurate but they are quite expensive computationally. To overcome the computational challenges especially for really large systems, single active electron (SAE) methods become advantageous, although some approximations are necessary in developing the model potentials \cite{Parker2000, Parker1998}. Reasonably accurate eigenvectors and energies can be generated using the model potentials.  The major limitation of SAE approximations is the inability to explain multiple electron features like double excitation, simultaneous excitation and ionization, double ionization, and innershell transitions. However, progress is being made towards the realization of these features. 
 
  The development of the single particle potentials is an active field of study taking different approximations \cite{Chang1976} like the independent particle approximation (IPA), multi-configurational Hartree-Fock (HF) \cite{Szabo1996}, density functional theory (DFT) \cite{Kohn1965}, random phase approximation (RPA) \cite{Linderberg1980}, and many others.
  
  Hartree \cite{Hartree1928}, Zener \cite{Zener1930} and Slater \cite{Slater1928, Slater1930a, Slater1930b} contributed immensely in the pioneering development and use of single particle potentials by adopting central screening potentials. The Hartree-Fock screening potentials are determined by using self-consistent variational methods \cite{Hartree1928}. Zener and Slater, on the other hand, respectively suggest the use of variationally \cite{Zener1930} and empirically determined \cite{Slater1928, Slater1930a, Slater1930b} atomic shielding constants which incorporate the effect of the other electrons in the effective nuclear charge on a single electron. The effective nuclear charge is then used in the solution of the Schr\"odinger equation for a hydrogenic system. The use of the central potentials is justified by the understanding that when more accurate energy levels are obtained, the wavefunctions obtained from the resulting boundary conditions will closely approximate the exact values when solving the atomic and molecular problems \cite{Slater1928}.  In this paper, we extend the concept of the effective nuclear charge further by suggesting a screening constant that depends on the local orbital angular momentum quantum number, $l$, of the single-particle electron and the nuclear charge, $Z$.
  
 In our previous works \cite{Jobunga2017b, Jobunga2017c, Jobunga2017e}, we have developed a theory for resolving the electron-electron interaction term with a goal of making the Hamiltonian separable. The separable Hamiltonian makes it possible to reduce the complex system to a one particle problem. The theory advanced requires the use of a suitable partition fraction for the results to be accurate. In reference \cite{Jobunga2017b, Jobunga2017c}, a classical partition fraction is suggested, but the method requires the use of an approximation to make the Hamiltonian separable. The classical partition function results into a central pseudopotential yielding reliable energies for excited states of $n-$electron atoms for $2\leq n \leq 12$, although it performs poorly for the groundstate energies. The equal partitioning of the electron-electron interaction energy, on the other hand, results into an exactly separable Hamiltonian. It is reasonably successful in predicting the non-relativistic groundstate energy for helium atom, within the lowest-order approximation adopted, due to its spherical symmetry in the groundstate. The equal partitioning, however, results into accidental degeneracy for states with the same principal quantum number and is quite poor in predicting energies for non-spherical states \cite{Jobunga2017b}.

 In this work, a symmetry-dependent partition fraction for helium atom is suggested. With the partition fraction, the corresponding lowest-order non-relativistic Hamiltonian is completely separable resulting into an independent particle problem. For spherically symmetric (${l=0}$) states, the equal partitioning is preserved with the suggested symmetry-dependent partition fraction. For non-spherical (${l\neq 0}$) states, the partition fraction is suggested to depend on the $l$-value so that the accidental degeneracy associated with equal partitioning is removed \cite{Jobunga2017b}. We obtain reliable results for the groundstate, excited states, and the doubly excited states of helium atom using the present method. 

\section{Theory}
The non-relativistic Hamiltonian, $H$, of a two-electron system with a nuclear charge $Z$, in atomic units, is given by
    \begin{equation}
    \mathrm{H} = \frac{1}{2}\, \left[p_1^2 + p_2^2\right] - Z\, \left[\frac{1}{r_1} + \frac{1}{r_2} \right] + \frac{1}{|\mathbf{r}_1-\mathbf{r}_2|}
    \end{equation}
where the first term correspond to the sum of the kinetic energies of each of the two electrons, the second term to the sum of the interactions between each of the electrons and the nucleus, and the last term to the electron-electron repulsion between the two electrons. The second and the last term form the potential energy function of a bound two-electron system.

 In our previous work \cite{Jobunga2017e}, it was shown that the electron-electron interaction simplifies to
 \begin{equation}
 \frac{1}{|\mathbf{r}_i-\mathbf{r}_j|} = \frac{1}{\sqrt{r_i^2 + r_j^2}} \label{eq:pt1c}
 \end{equation}
in the lowest-order approximation. This is in agreement with the formulation used in the hyperspherical method \cite{Macek1968}. In the independent particle method, the single-electron potential function 
 \begin{equation}
 V(r_i, r_j) = -\frac{Z}{r_i} + \eta_{l_i}\,\frac{1}{\sqrt{r_i^2 + r_j^2}} \label{eq:pt1}
 \end{equation}
for the $i^{\mathrm{th}}-$electron in a two-electron system can be completely separated \cite{Jobunga2017b} as
\begin{equation}
 V(r_i) = -\frac{Z- \sigma}{ r_i}. \label{eq:pt1b}
 \end{equation}
 where
 \begin{equation}
 \sigma = \eta_{l_i}\,\sqrt[3]{\frac{Z}{\eta_{l_i}}}
 \end{equation}
 is the screening (or shielding) constant emanating from the effect of the other electron. The separarability is achieved by imposing the condition
 \begin{equation}
 \frac{\partial V(r_i,r_j)}{\partial r_i} = 0
 \end{equation}
 which ensures that the potential energy function in Eq. \eqref{eq:pt1} is minimum.
 
 From Eq. \eqref{eq:pt1b}, it can be deduced that the nuclear charge screening parameter due to one electron on the other electron can be determined exactly within the lowest-order approximation. The charge screening can be seen to depend on the nuclear charge and the angular momentum of the active electron. This work therefore modifies the existing theory of charge screening \cite{Clementi1963} by introducing symmetry dependence in it.  
 
 Factor $\eta_{l_i}$ in equations (\ref{eq:pt1}) and (\ref{eq:pt1b}) corresponds to a partition fraction which ensures the sharing of the electron-electron interaction energy between the two interacting electrons as a function the orbital angular momentum of the $i^{\mathrm{th}}$ electron. We have seen in our previous work \cite{Jobunga2017b, Jobunga2017c} that the equal sharing (${\eta_{l_i}=1/2}$) of the electron-electron interaction is  reasonably successful in approximating, to the lowest-order, the groundstate ionization potential of helium atom because of its spherical symmetry.
 
 In this work, a symmetry-dependent partition fraction 
 \begin{equation}
 \eta_{l_i} = \frac{ 1 + \delta_{l_i}}{ 2 + \delta_{l_i}} \label{eq:pt2}
 \end{equation}
  for helium atom, with the parameter
  \begin{equation}
  \delta_{l_i}= \sqrt[l_i]{l_i} \label{eq:pt2b}
  \end{equation}
empirically determined, is suggested. With the partition fraction, the corresponding lowest-order non-relativistic Hamiltonian is completely separable leading to a single-electron problem. The partitioning can be explained on the basis that the two electrons with share the electron-electron interaction energy as a fraction of their intrinsic energies. The electrons are assumed to be quantum harmonic oscilators whose intrinsic energies are given by
\begin{equation}
\epsilon_i = \left[\tilde{l}+\frac{1}{2}\right]\hbar \omega
\end{equation} 
 where  $\tilde{l}$ is a discrete quantum number, $\hbar$ is the Planck's constant divided by ${2\pi}$, and $\omega$ is the angular frequency. With this argument, the partition fraction can be seen to take the form
 \begin{equation}
 \eta = \frac{\epsilon_i}{\epsilon_i + \epsilon_j}  = \frac{ 1 + \delta_{l_i}}{ 2 + \delta_{l_i} + \delta_{l_j}}
 \end{equation}
where $\delta_{l_i}$ and $\delta_{l_j}$ are local and non-local symmetry ($l$) dependent terms respectively. For helium atom, the non-local term is assumed to be zero and the local term is empirically determined as indicated in Eq. \eqref{eq:pt2b}.
 
With the saparable potential given by Eq. \eqref{eq:pt1b}, the single-electron Hamiltonian, 
 \begin{equation}
 h(r_i)= \frac{p_i^2}{2} + V(r_i), \label{eq:pt3}
 \end{equation}
for helium atom is defined.  The eigenvalues of a two-electron system can then be evaluated as \cite{Jobunga2017b}
 \begin{equation}
  \langle E_{\alpha \alpha'} \rangle  = \left\{ \begin{matrix}
  4\, {\varepsilon}_{\alpha \alpha'} & \mathrm{if}\; \alpha = \alpha'\\
  {\varepsilon}_{\alpha \alpha} + {\varepsilon}_{\alpha' \alpha'} & \mathrm{if}\; \alpha \neq \alpha'
  \end{matrix} \right. \label{eq:pt4}
 \end{equation}
 where ${{\varepsilon}_{\alpha \alpha} = \langle h(r_i) \rangle }$ is the eigenvalue of a single electron orbital.  Factor $4$ in Eq. \eqref{eq:pt4} arises from both exchange and permutation symmetry consideration for states with $\alpha=\alpha'$. For other cases where ${\alpha \neq \alpha'}$, the inner electron sees the unscreened nuclear charge $Z$ while the outer electron sees the screened nuclear charge as a result of the inner electron. The energy eigenvalue, corresponding to the principal quantum number $n$, for the inner electron can therefore be computed as a hydrogenic eigenvalue 
 \begin{equation}
 \varepsilon_n=-\frac{Z_{eff}^2}{2n^2} \label{eq:pt5}
 \end{equation}
  with unscreened nuclear charge ${Z_{eff}=Z}$ whereas that for the outer electron can also be computed in a similar way but with a screened nuclear charge ${Z_{eff}=Z-\sigma}$. For a helium atom with one electron considered to be in the $1s$ state and the other electron occupying an excited state $\alpha'$, ${\varepsilon}_{1s}$  is equal to the energy eigenvalue, ${E_{\mathrm{1s}}=-2.00000}$, for the helium ion in its ground state since the inner electron is unshielded.
 
\section{Results and Discussions}

We have developed a single electron potential for helium atom given by equation (\ref{eq:pt1b}). The pseudopotential is used to calculate the $1snl$ eigenvalues for helium atom as shown in table \ref{tab1} for angular momenta of up to ${l_{\mathrm{max}}=7}$. The results, calculated using Eq.\eqref{eq:pt5}, are presented for the first five principal quantum numbers for each angular momentum value. 

\begin{table}[!ht]
    \centering
    \begin{tabular}{m{1.8cm}m{1.8cm}m{1.8cm}m{1.8cm}m{1.8cm}}
    \hline
    State & Present & Trip. & Sing.  & Ref.  \\
   \hline
   \hline
       $L=0$&-2.91031   & -        &-2.90394 & -2.90372   \\                                                     
            &-2.18189   &-2.17528  &-2.14601 & -2.14597   \\                          
            &-2.08084   &-2.06871  &-2.06128 & -2.06127  \\  
            &-2.04547   &-2.03652  &-2.03359 & -2.03358   \\ 
            &-2.02910   &-2.02262  &-2.02118 &    \\                                  
    \hline
       $L=1$&-2.13481   &-2.13320  &-2.12387 &  -2.12384  \\                                                     
            &-2.05991   &-2.05809  &-2.05516 &  -2.05514  \\                          
            &-2.03370   &-2.03233  &-2.03107 &  -2.03106  \\  
            &-2.02156   &-2.02055  &-2.01991 &  -2.01991  \\ 
            &-2.01497   &-2.01421  &-2.01383 &    \\                                  
    \hline
       $L=2$&-2.05555   &-2.05565  &-2.05563 &  -2.05562  \\                                                     
            &-2.03124   &-2.03129  &-2.03128 &  -2.03127  \\                          
            &-2.01999   &-2.02002  &-2.02002 &  -2.02001  \\  
            &-2.01388   &-2.01390  &-2.01390 &  -2.01389  \\ 
            &-2.01020   &-2.01021  &-2.01021 &    \\                                  
    \hline
       $L=3$&-2.03110   &-2.03126  &-2.03126 &   -2.03125 \\                                                     
            &-2.01991   &-2.02000  &-2.02000 &   -2.02000 \\                          
            &-2.01382   &-2.01389  &-2.01389 &   -2.01389 \\  
            &-2.01015   &-2.01020  &-2.01020 &   -2.01020 \\ 
            &-2.00777   &-2.00781  &-2.00781 &    \\                                  
    \hline
       $L=4$&-2.01999   &-2.02000  &-2.02000 &   -2.02000 \\                                                     
            &-2.01388   &-2.01389  &-2.01389 &   -2.01388\\                          
            &-2.01020   &-2.01020  &-2.01020 &   -2.01020 \\  
            &-2.00781   &-2.00781  &-2.00781 &    \\ 
            &-2.00617   &-2.00617  &-2.00617 &    \\                                  
    \hline
       $L=5$&-2.01396   &-2.01389  &-2.01389 &   -2.01388 \\                                                     
            &-2.01026   &-2.01020  &-2.01020 &   -2.01020 \\                          
            &-2.00785   &-2.00781  &-2.00781 &   -2.00781 \\  
            &-2.00620   &-2.00617  &-2.00617 &    \\ 
            &-2.00502   &-2.00500  &-2.00500 &    \\                                  
    \hline
       $L=6$&-2.01031   &-2.01020  &-2.01020 &   -2.01020 \\                                                     
            &-2.00789   &-2.00781  &-2.00781 &   -2.00781 \\                          
            &-2.00624   &-2.00617  &-2.00617 &   -2.00617 \\  
            &-2.00505   &-2.00500  &-2.00500 &    \\ 
            &-2.00417   &  & &    \\                                  
    \hline
       $L=7$&-2.00793   &-2.00781  &-2.00781 &   -2.00781 \\                                                     
            &-2.00626   &-2.00617  &-2.00617 &   -2.00617 \\                          
            &-2.00507   &-2.00500  &-2.00500 &   -2.00499 \\  
            &-2.00419   &  & &    \\ 
            &-2.00352   &  & &    \\                                  
    \hline
    \end{tabular}
    \caption{Some numerically calculated eigenvalues using the present method potential versus the experimentally determined triplet and singlet values \cite{Nist2018} and the non-relativistic reference values for helium atom \cite{Scrinzi1998}}.
    \label{tab1}
  \end{table}

The results generated with the derived pseudopotential are in reasonable agreement with the references values \cite{Scrinzi1998} at larger values of $n$ and $l$ as expected. The singlet and triplet values presented in table \ref{tab1} are the reference experimental \cite{Nist2018} results. We can observe that the present results are close to the triplet values while the reference theoretical data \cite{Scrinzi1998} are close to the singlet values. It is also clearly visible in the present results that the accidental degeneracy observed in our previous results \cite{Jobunga2017b}, where states having the same principal quantum numbers but different angular momentum quantum numbers have the same energies, is completely removed. This is a consequence of the symmetry-dependent interaction potential used in our suggested method.

The largest discrepancy between the present results and the literature values are in the lowest lying spherically symmetric states. The discrepancy seems to stem from the equal partitioning of the electron-electron interaction for the ${l=0}$ states. In ref. \cite{Jobunga2017b}, it is evident that the classical partitioning of the electron-electron interaction, which is dependent on the radial distance $r_i$, yields better aggreement with the excited spherical states.

In table \ref{tab2}, we present the eigenvalues of autoionizing levels of helium relative to the groundstate energy. The eigenvalues have been evaluated using equation (\ref{eq:pt4}).   
\begin{table}[!ht]
    \centering
    \begin{tabular}{m{1.8cm}m{1.8cm}m{1.8cm}m{1.8cm}}
    \hline
    State & Present & ref$_1$ & ref$_2$  \\
   \hline
   \hline
       $2s^2$ $ ^1S$& -0.7275   &  -0.7333 & -0.7778    \\                                   
%
     $3s^2$ $ ^1S$& -0.3233      & -0.3265 & -0.3535    \\                          
%
    $4s^2$ $ ^1S$&   -0.1818    & -0.1838  & -0.2010   \\ 
    $5s^2$ $ ^1S$&   -0.1164    & -0.1177  & -0.1303   \\
    $6s^2$ $ ^1S$&   -0.0808    & -0.0817  & -0.0908 \\
    $7s^2$ $ ^1S$&   -0.0593    & -0.0600  & -0.0675  \\
    $2s2p$ $ ^1P$&   -0.6819    & -0.6587  & -0.6931  \\
    $2p^2$ $ ^1S$&   -0.5392    & -0.6314  & -0.6219  \\
    $3p^2$ $ ^1S$&   -0.2396    & -0.2933  & -0.3174  \\
    $4p^2$ $ ^1S$&   -0.1348    & -0.1671  & -0.1832  \\
    $5p^2$ $ ^1S$&   -0.0862    & -0.1075  & -0.1210  \\
    $6p^2$ $ ^1S$&   -0.0599    & -0.0749  & -0.0857   \\
    $7p^2$ $ ^1S$&   -0.0440    & -0.0551  & -0.0641   \\               
    \hline
    \end{tabular}
    \caption{Some numerically calculated eigenvalues (in atomic units) using the present potential versus the literature values\cite{Ranbir1996} for helium autoionizing levels. Ref.$_1$ results were generated using DFT calculations while ref.$_2$ are reference data reported in the paper for comparison purposes.}
    \label{tab2}
  \end{table}

From table \ref{tab2}, one can observe that our $ns^2$ $^1S$ states and $2s2p$ $^1P$ state results compare well with ref.$_1$ values while $np^2$ $^1S$ states results are lower compared to the reference values. This observation forms a complementary relationship with table \ref{tab1} results in that singly excited states with $s$ symmetry are poorly described while doubly excited states with $s$ symmetry are well described by the present potential. Conversely, single excited states with $p$ symmetry are well described while doubly excited states with $p$ symmetry are underestimated by the present potential. 

The non-relativistic excitation energies of various autoionizing states, presented in table \ref{tab3}, have also been determined using the derived pseudopotential. The results are compared with known experimental results \cite{Rudd1964}.

\begin{table}[!ht]
    \centering
    \begin{tabular}{m{1.8cm}m{1.8cm}m{1.8cm}m{1.8cm}m{1.8cm}m{1.8cm}}
    \hline
    State & Present (eV) & Sing. & Exp. (Sing.) & Trip. & Exp. (Trip.) \\
   \hline
   \hline
       $2s^2$ & 59.19   &  57.95 & 57.8 & &    \\                                   
%
     $2s2p$ & 60.43     & 60.13 & 60.1 & 58.22 & 58.3 \\                          
%
    $2s3s$ &   63.18    & 63.09  & 62.9 &62.68 &  \\ 
    $2s4s$ &   64.14    & 64.25  &  & 64.11&  \\
    $2s5s$ &   64.59    & 64.71  &  & 64.65 &\\
    $2p^2$ &   64.31    & 63.00  & 62.2 & & \\
    $2p3p$ &   63.75    & 64.28  &  & 63.85& \\
    $2p3d$ &   63.87    & 64.19  &  & 64.21& \\
    $2p4p$ &   64.97    & 64.73  &  & 64.55& \\
    $2p4d$ &   64.53    & 64.69  &  & 64.70& \\
    \hline
    \end{tabular}
    \caption{Some numerically calculated excitation energies (in eV) for the autoionizing states of helium atom using the present pseudpotential versus the singlet (Sing.) and triplet (Trip.) theoretical \cite{Macek1968} and experimental \cite{Rudd1964} literature values.}
    \label{tab3}
  \end{table}

The excitation energies calculated using the present method are in fair agreement with the literature values except for the $2s^2$ and $2p^2$ states where our present results are higher because the exchange effects. It is important to note that the present method can neither resolve the singlet and triplet states of the autoionizing levels nor the $S$ and $D$ states for the $npn'p$ autoionizing levels.

The present method is advantageous in that, unlike other methods tackling the electron correlation effects, there is no use of self consistent approximations or iterations involved in the treatment of the electron-electron interaction and all the effects including electron exchange and correlation are evaluated exactly within the lowest-order approximation given by Eq.\eqref{eq:pt1c}. The validity of the current method lies in the suggested partition function which in turn determines the quality of the results obtained. 

Within the non-relativistic solution framework for helium atom, the present results are in good comparison with the literature values and can be deemed  to be reliable, especially for non-spherical ($l \neq 0$) symmetry states. For states with spherical ($l=0$) symmetry, the method is only accurate for the groundstate and autoionizing states while it overstimates the binding energies of the singly excited states. 

With the symmetry-dependent partition fraction, the separation of the correlated electron-electron interaction is exactly solved within the lowest-order approximation. For helium atom, we are led to a conclusion that the lowest-order non-relativistic ground state energy is $-2.9103$. This implies that the accurate value of $-2.9037$ can only be achieved if the higher-order non-relativistic corrections are incorporated into the Hamiltonian of helium atom. This premise will be a subject of further investigation in future.

\section{Conclusion}
Accurate treatment of the electron-electron interation is the key to resolving uncertainties in many-body physics. Many existing methods for solving the many-body problems are quite expensive computationally. In this work, a symmetry-dependent partition function for helium atom is suggested. A simple separable parameter-free Hamiltonian for helium atom yielding reasonable and degenerate non-relativistic eigenvalues is consequently obtained. The problem reduces to an effective potential approach with a charge screening parameter exactly known within the lowest-order approximation. The singly and doubly excited state energies obtained by the separable potential compare well with literature values. It is hoped that the suggested method will go along way in improving the solutions of the complex multi-electron problems.

\section{Acknowledgement}
We are grateful to the German Centre for Migration and Development (CIM) for donating the computational facilities used in this research. 

\bibliographystyle{apsrev}
\bibliography{/home/eric/Inworks/Literature}

\end{document}